\documentclass[prd,preprint]{revtex4}
\usepackage[dvips]{graphicx}
\usepackage{url}

\begin{document}

\title{Classical QGP : IV. Thermodynamics}

\author{Sungtae Cho and Ismail Zahed}
\email{scho@grad.physics.sunysb.edu,
zahed@zahed.physics.sunysb.edu}
\address{Department of Physics and Astronomy \\
State University of New York, Stony Brook, NY, 11794}

\begin{abstract}
We construct the equation of a state of the classical QGP valid
for all values of $\Gamma=V/K$, the ratio of the mean Coulomb to
kinetic energy. By enforcing the Gibbs relations, we derive the
pertinent pressure and entropy densities for all $\Gamma$. For the
case of an SU(2) classical gluonic plasma our results compare well
with lattice simulations. We show that the strongly coupled component
of the classical QGP contributes significantly to the bulk thermodynamics
across $T_c$.
\end{abstract}

\maketitle

\newpage

\section{Introduction}

Classical Plasmas are statistical systems with constituents that
are locally charged but globally neutral. An example is the
electromagnetic one component plasma (OCP) also referred to as
Jelium. A number of many body theories have been devised to
analyze the OCP in the regime of small $\Gamma=V/K$, the ratio of
the average Coulomb energy to kinetic energy~\cite{ichimaru}. Most
of the extensions to larger values of $\Gamma$ are based on higher
order transport equations~\cite{hansen&mcdonald} or classical
molecular dynamics~\cite{hansenetal}.

The Classical Quark Gluon Plasma (cQGP) as developed by Gelman,
Shuryak and Zahed can be regarded as an extension of the OCP
plasma to many components with non-Abelian color charges
\cite{borisetal}. Stability against core collapse is enforced
classically through a phenomenological core potential. The origin
of the core is quantum mechanical. Detailed molecular dynamics
simulations of the cQGP~\cite{borisetal} have shown a strongly
coupled plasma for $\Gamma=V/K\approx 1$ or larger. The cQGP
maybe in a liquid state at moderate values of $\Gamma$. In a
recent analysis~\cite{cho&zahed} we have used analytical methods
of classical liquids to construct the free energy for small
$\Gamma$ both in the dilute case and at high temperature after
resummation of the screening effects.

In this paper we combine the results in~\cite{borisetal} obtained
from molecular dynamics with the one-loop analytical results
in~\cite{cho&zahed} to construct the equation of state of the cQGP
for all values of $\Gamma$. We will show that the strongly coupled
component of the cQGP contributes significantly to the
thermodynamics across the transition whether in the energy
density, pressure or entropy density. In section 2, we derive the
excess energy of the cQGP for small $\Gamma$ in the one-loop
approximation. In section 3, we use ideas from classical
electromagnetic plasmas to interpolate between the one-loop result
at low $\Gamma$ and the molecular dynamics results at large
$\Gamma$ for the SU(2) cQGP. Particular attention will be given to
the core parameter using Debye-H\"uckel plus Hole (DHH) theory. In
the quantum QCD plasma, $\Gamma$ runs with $T$. In section 4, we
use the interpolated excess energy density together with the Gibbs
relations to derive the pressure and entropy of the cQGP. In
section 5, we compare the results for the SU(2) cQGP with
SU(2) lattice simulations. Our conclusions are in section 6. In
Appendix A we show that the interaction corrections to the
concentration do not affect the one-loop result to order
$\Gamma^{\frac 52}$ with the bare particle concentration. In Appendix B
we summarize the Debye-H\"uckel plus Hole theory to assess the
range of the core in terms of the Debye length.

\section{Excess energy: One-Loop}

\renewcommand{\theequation}{II.\arabic{equation}}
\setcounter{equation}{0}

In classical plasmas the key expansion parameter at zero
chemical potential is $\Gamma=V/K$ the ratio of the mean
Coulomb to kinetic energy. For an Abelian or QED plasma,

\begin{equation}
\Gamma=\frac{(Ze)^2}{a k_B T} \label{eq001f}
\end{equation}
while for a non-Abelian or QCD plasma~\cite{borisetal}

\begin{equation}
\Gamma=\frac{g^2}{4\pi}\frac{C_{2}}{T a_{WS}} \label{eq002f}
\end{equation}
with $k_B=1$ and $a_{WS}$ the Wigner-Seitz
radius satisfying $N/V(4\pi a_{WS}^3/3)=1$. $C_2$ is the quadratic
Casimir ($C_2=q_2/(N_c^2-1))$ and $g$ is the strength of the
coupling. In the cQGP $g$ is fixed, while in the QGP $g$ runs
with temperature. The running is quantum mechanical and beyond
the present classical analysis. In sections 4,5 it will be addressed
phenomenologically.

Since the Wigner-Seitz radius $a_{WS}$ is tied with the density or
concentration (the bare concentration is $c_0=N/V$), it is
straightforward to express the free energy in terms of $\Gamma$.
After resumming the screening effects and to one-loop, the free
energy reads~\cite{cho&zahed}

\begin{eqnarray}
\beta\frac{F_{loop}(\beta,c)}{V}&=&-c-\frac{2\sqrt{\pi}}{3}(N_c^2-1)
\gamma^{\frac{3}{2}}c^{\frac{3}{2}}+\pi(N_c^2-1)c^2\gamma^2\sigma
\nonumber\\ &&-2\pi^{\frac{3}{2}}(N_c^2-1)c^{\frac{5}{2}}\gamma^{\frac{5}{2}}
\sigma^2 + c\beta\mu_c+\mathcal{O}(\beta^3) \label{eq003f}
\end{eqnarray}
with $\gamma=g^2/4\pi\beta C_2$, $c$ the concentration and
$\sigma$ the core radius. The concentration is determined by the
chemical equation~\cite{cho&zahed} and to leading order is
$c_0=N/V$ as detailed in Appendix A. The core $\sigma$ is a
parameter in the cQGP much like in normal classical liquids. Its
origin is quantum mechanical. In Appendix B, we use the classical
Debye-H\"uckel plus Hole (DHH) theory to assess the size of the
core in terms of the Debye screening length.

To see how the expansion in the concentration $c$ in
(\ref{eq003f}) converts to an expansion in $\Gamma$, we note that
the Debye-H\"uckel contributions (first two terms) can be
rewritten as

\begin{eqnarray}
\frac{F_{DH}(\Gamma)}{NT}&=&-c\frac{V}{N}-c^{\frac{3}{2}}
\frac{V}{N}\frac{2\sqrt{\pi}}{3}(\frac{g^2}{4\pi})^{\frac{3}{2}}
(N_c^2-1)(\beta C_{2})^{\frac{3}{2}} \nonumber \\
&=&-1-c^{\frac{3}{2}}\frac{4\pi}{3}a_{WS}^3\frac{(4\pi)^{\frac{1}
{2}}}{3} (\frac{g^2}{4\pi})^{\frac{3}{2}}(N_c^2-1)(\beta
C_{2})^{\frac{3}{2}} \nonumber \\
&=&-1-\frac{1}{\sqrt{3}}c^{\frac{3}{2}} \Big( \frac{4\pi
a_{WS}^3}{3}\Big)^{\frac{3}{2}} (N_c^2-1) (\frac{g^2}{4\pi}\beta
\frac{C_{2}}{a_{WS}})^{\frac{3}{2}} \nonumber \\
&=& -1-\frac{1}{\sqrt{3}} (N_c^2-1) \Gamma^{\frac{3}{2}}
\label{eq004f}
\end{eqnarray}
By defining the excess free energy $F_{ex}$ as $F_{ex}(\Gamma)= F(\Gamma)-F(0)$
we obtain to one-loop,

\begin{equation}
\frac{F_{loop,ex}}{NT}=-\frac{1}{\sqrt{3}}(N_c^2-1)\Gamma^{\frac{3}{2}}
+\frac{3}{4}\delta(N_c^2-1)\Gamma^2-3\sqrt{3}\delta^2(N_c^2-1)
\Gamma^{\frac{5}{2}}+\mathcal{O}(\Gamma^3) \label{eq005f}
\end{equation}
with $\delta=\sigma/a_{WS}$. $F(0)$ will be identified with the
free gas or Stephan-Boltzman contribution.

The excess energy $U_{ex}$ of the cQGP follows from the excess free
energy $F_{ex}$ as

\begin{equation}
\frac{F_{ex}(\Gamma)}{NT}=\int_{0}^{\Gamma}\frac{U_{ex}}{NT}
\frac{d\Gamma'}{\Gamma'}
\label{eq006f}
\end{equation}
For instance, the Debye-H\"uckel contribution in (\ref{eq007f})
yields through (\ref{eq006f}) the excess energy

\begin{equation}
\frac{U_{DH,ex}}{NT}=-
\frac{\sqrt{3}}{2}(N_c^2-1)\Gamma^{\frac{3}{2}} \label{eq007f}
\end{equation}
in agreement with the Debye-H\"uckel excess energy for the cQGP
initially discussed in~\cite{borisetal2} using different methods.
In general, the energy density splits into the free plus excess
the contribution

\begin{equation}
\epsilon(\Gamma)=\frac{U(\Gamma)}{V}=\frac{U_0}{V}+\frac{U_{ex}(\Gamma)}{V}
=\epsilon_0+\epsilon(\Gamma) \label{eq008f}
\end{equation}
with the free contribution $\epsilon_0=\epsilon_{SB}$ identified with
the Stefan-Boltzmann energy density $\epsilon_{SB}$. In relative
notations,

\begin{equation}
\frac{\epsilon(\Gamma)}{\epsilon_{SB}}=1+\frac{1}
{\epsilon_{SB}} \frac{U_{ex}(\Gamma)}{V}  \label{eq009f}
\end{equation}
Using (\ref{eq006f}) together with (\ref{eq005f}), we obtain
the one-loop excess energy density

\begin{equation}
\frac{U_{loop,ex}}{NT}=-\frac{\sqrt{3}}{2}(N_c^2-1)\Gamma^{\frac{3}{2}}
+\frac{3}{2}\delta(N_c^2-1)\Gamma^2-\frac{15}{2}\sqrt{3}\delta^2(N_c^2-1)
\Gamma^{\frac{5}{2}}+\mathcal{O}(\Gamma^3) \label{eq014f}
\end{equation}
which is valid for small $\Gamma$. In Appendix A, we show that although
the concentration $c$ in $\Gamma$ is not $c_0$ because of interactions,
to order $\Gamma^{5/2}$ we may set $c=c_0$.

\section{Excess Energy: Full}

\renewcommand{\theequation}{III.\arabic{equation}}
\setcounter{equation}{0}

In~\cite{cho&zahed} the one-loop expansion was shown to converge
up to $\Gamma\approx 1$ for the free energy. The range is even smaller
for the energy with $\Gamma\approx 1/2$ (see below). Larger values of
$\Gamma$ have been covered by molecular dynamics simulations in~\cite{borisetal}.
For an SU(2) plasma (say a constituent gluonic plasma) the numerical results
for the excess energy were found to follow the parametric form
~\cite{borisetal}

\begin{equation}
\frac{U_{mol}}{NT}\simeq-4.9-2\Gamma+3.2\Gamma^{\frac{1}{4}}+\frac{2.2}
{\Gamma^{\frac{1}{4}}}\,\,.
\label{eq001s}
\end{equation}
For $N_c=2$ the one-loop result (\ref{eq014f}) reads

\begin{equation}
\frac{U_{anal}}{NT}=-\frac{3}{2}\sqrt{3}\Gamma^{\frac{3}{2}}
+\frac{9}{2}\delta\Gamma^2-\frac{45}{2}\sqrt{3}\delta^2\Gamma^{\frac{5}{2}}
\label{eq002s}
\end{equation}
To construct the full excess energy valid for all $\Gamma$ we will
proceed phenomenologically by seeking an interpolating formulae
between (\ref{eq001s}) and (\ref{eq002s}) borrowing from ideas in
classical plasma physics~\cite{ichimaru2}. A similar approach was
also advocated in \cite{bannur} using different limits.

In the Abelian or QED plasma, the excess energy based on
Debye-H\"uckel theory is evaluated for $\Gamma<0.1$. Molecular
dynamics simulations are generated for $1<\Gamma<180$. The two are
combined numerically through a power function in the
form~\cite{ichimaru2}

\begin{figure}[!h]
\begin{center}
\includegraphics[width=0.50\textwidth]{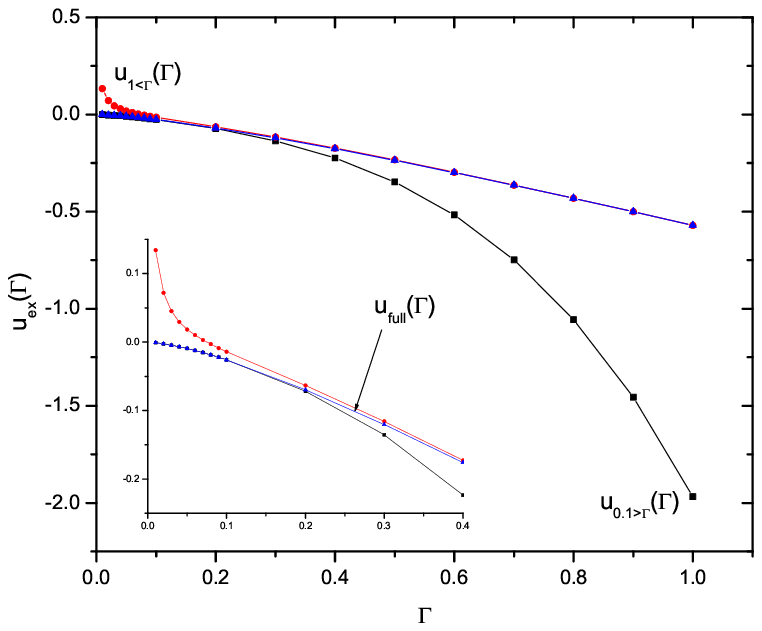}
\includegraphics[width=0.49\textwidth]{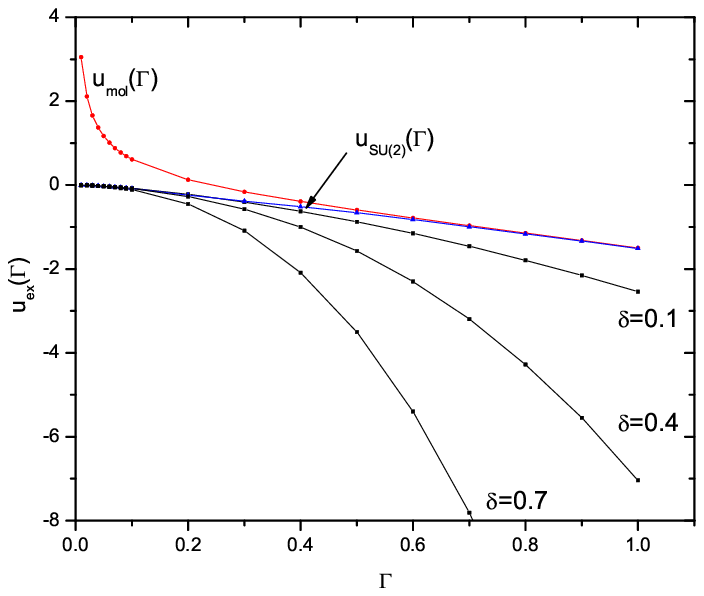}
\end{center}
\caption{Excess energy for QED (left) and $N_c=2$ QCD (right)}
\label{connected}
\end{figure}

\begin{equation}
u_{full}(\Gamma)=\frac{u_{\Gamma<0.1}(\Gamma)+f(\Gamma)u_{\Gamma>1}(\Gamma)}
{1+f(\Gamma)} \label{eq003s}
\end{equation}
with $f(\Gamma)$ a fitting power function of the type
$a\Gamma^b(=3.0\times10^3\Gamma^{5.7})$. (\ref{eq003s})
interpolates smoothly between the exact analytical results
at low $\Gamma$ and the simulations at large $\Gamma$ as
shown in Fig.~\ref{connected} (left). In the insert we
show the nature of the size of the gap in the range
$0.1<\Gamma<1$ for the Abelian plasma.

In Fig.~\ref{connected} (right) we show our $N_c=2$ results
at low values of $\Gamma$ (one-loop) and large values of
$\Gamma$ (simulations). The one-loop results depend on the
size of the hard core $\sigma$. Recall that the simulations
in ~\cite{borisetal} are carried with a fixed higher power law
repulsion to mock up the core. So
the simulations seem to favor a small hard core
in the Wigner-Seitz units. In fact, $\sigma$ can be set by the Debye
radius in the DHH theory~\cite{nordholm} detailed in Appendix B.
It changes with $\Gamma$. Specifically,

\begin{figure}[!h]
\begin{center}
\includegraphics[width=0.50\textwidth]{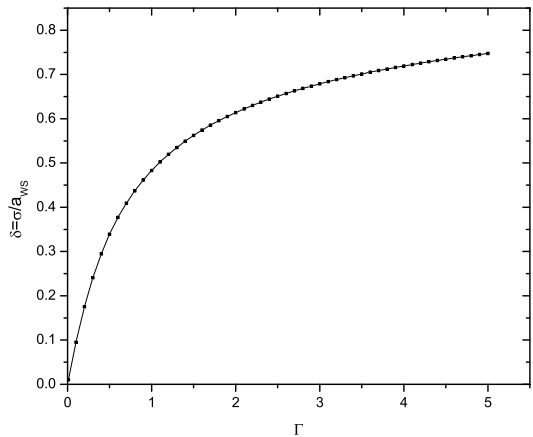}
\end{center}
\caption{Core parameter $\delta=\sigma/a_{WS}$. See text.}
\label{delta}
\end{figure}

\begin{equation}
\delta=\frac{\sigma}{a_{WS}}=\frac{1}{a_{WS}\kappa_{D}}\Big((1+(3\Gamma)
^{\frac{3}{2}})^{\frac{1}{3}}
-1\Big)=\frac{1}{(3\Gamma)^{\frac{1}{2}}}\Big((1+(3\Gamma)
^{\frac{3}{2}})^{\frac{1}{3}} -1\Big) \label{eq004s}
\end{equation}
which is shown in Fig.~\ref{delta}. The core size $\delta$ varies
in the range $0.2-0.5$ for $\Gamma$ in the range $0.1-1$. We fix
$\delta=0.4$ in the range $0.1-1$. With this in mind and following
the Abelian plasma construction, we find the excess energies shown
in Fig.~\ref{connected} (right) to be fit by

\begin{equation}
u_{SU(2)}(\Gamma)=\frac{u_{anal}(\Gamma)+5.5\times10^2\Gamma^{5.4}
u_{mol}(\Gamma)} {1+5.5\times10^2\Gamma^{5.4}} \label{eq005s}
\end{equation}
The power function is numerically adopted to yield a small deviation
(less than $0.1\%$) for $\Gamma<0.1$ and $\Gamma>1$. The precise
choice of the core parameter is actually not very important, as small
changes in core size can be compensated by small changes in the power
function for the same overall accuracy.

\section{Thermodynamics}

\renewcommand{\theequation}{IV.\arabic{equation}}
\setcounter{equation}{0}

Knowledge of the energy density for all values
of $\Gamma$ can be used to extract all extensive
thermodynamical quantities in the cQGP with the help of the
Gibbs relations. Indeed, the pressure and entropy
follow from the Gibbs relations

\begin{eqnarray}
&&\epsilon=T\frac{\partial{P}}{\partial{T}}-P\nonumber\\
&&s=\frac SV=\frac 1V \frac{\partial P}{\partial T}\,\,.
\label{eq010f}
\end{eqnarray}

In so far the classical plasma parameter $\Gamma$ as defined
(\ref{eq002f}) is a fixed parameter. However, in QCD it runs
through $\alpha_s$. It is only a function of temperature.
Specifically,

\begin{equation}
\Gamma=\frac{\alpha_s C_{2}}{T a_{WS}}=\Big(\frac{4\pi}{3}c_0
\Big)^{\frac{1}{3}}\beta C_2\alpha_s(T)=\Big(0.244(N_c^2-1)
\frac{4\pi}{3}\Big)^{\frac{1}{3}}C_{2}\alpha_s(T)
\label{eq012f}
\end{equation}
where $0.244 (N_c^2-1)$ is the black-body concentration for
adjoint gluons. The exact running of $\alpha_s(T)$ will be
fixed below.

Using (\ref{eq010f}) together with (\ref{eq012f}) yield the
pressure and the entropy density directly in terms of the
energy density

\begin{eqnarray}
&&\frac{P}{P_{SB}}=3\frac{1}{T^3}
\int_{T_c}^{T}dT'{T'}^2\frac{\epsilon}{\epsilon_{SB}}(T')\nonumber\\
&&\frac{s}{s_{SB}}=
\frac{3}{4}\frac{\epsilon}{\epsilon_{SB}}(T)+\frac{3}{4}\frac{1}{T^3}
\int_{T_c}^{T}dT'{T'}^2\frac{\epsilon}{\epsilon_{SB}}(T')\,\,.
\label{eq013f}
\end{eqnarray}
Here $T_c$ is identified with the SU(2) transition with
$P_c=0$. For a constituent gluonic plasma $T_c=215$ MeV.
All bulk thermodynamics is tied to the energy density by the Gibbs
relations.

\section{SU(2) Lattice Comparison}

\renewcommand{\theequation}{V.\arabic{equation}}
\setcounter{equation}{0}

\begin{figure}[!h]
\begin{center}
\includegraphics[width=0.55\textwidth]{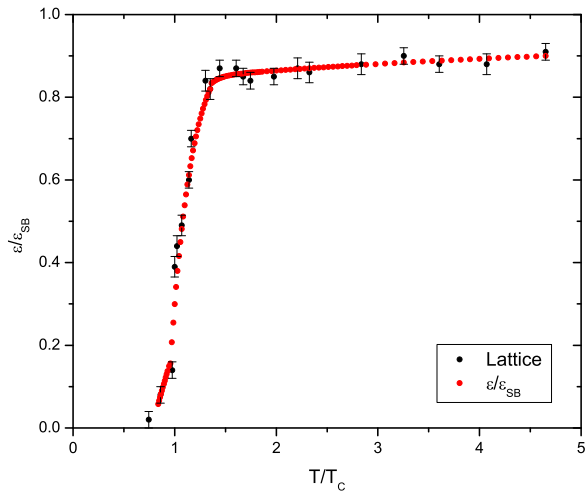}
\end{center}
\caption{Energy density fit to SU(2) lattice data. See text.}
\label{energy-lattice}
\end{figure}

To proceed further we need to know how $\alpha_s(T)$ runs with $T$
in pure YM and QCD, to determine the behavior of the extensive thermodynamical
quantities. The loop expansion allows a specific determination of
the running $\alpha_s(T)$ that is unfortunately valid at high
temperature or weak coupling. How $\alpha_s(T)$ runs at strong
coupling is unknown. Here we suggest to {\it extract}
$\alpha_s(T)$ from the lattice data by {\it fitting} our energy
density (\ref{eq009f}) valid for all couplings to the SU(2)
lattice data  in~\cite{engelsetal}. In Fig.~\ref{energy-lattice}
we show the fit of the normalized energy density in the cQGP
to the SU(2) lattice data. The lattice results are used with
$\Lambda_L=5$ MeV and $T_c=215$ MeV as suggested in~\cite{engelsetal}.

\begin{figure}[!h]
\begin{center}
\includegraphics[width=0.50\textwidth]{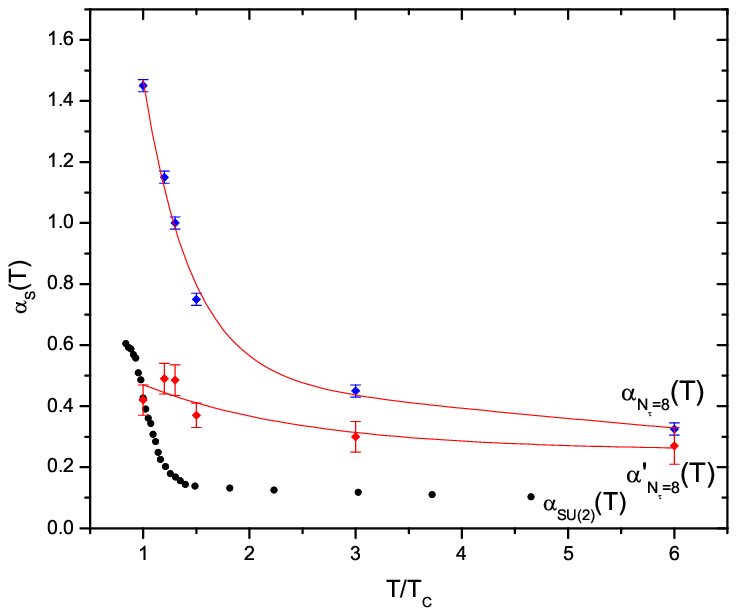}
\includegraphics[width=0.49\textwidth]{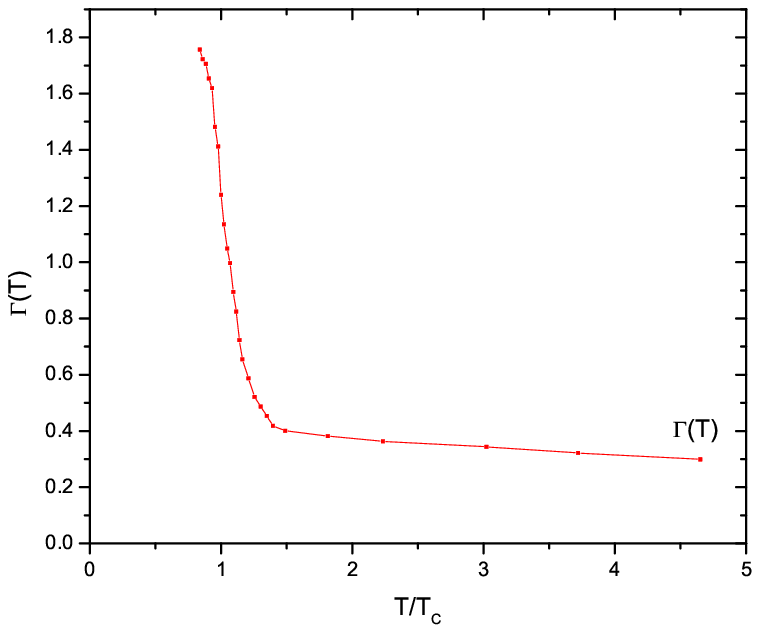}
\end{center}
\caption{$\alpha_s(T)$ (left) and $\Gamma (T)$ (right) for $N_c=2$. See text.}
\label{su2alpha}
\end{figure}

The energy density fit in return yields through
(\ref{eq009f}), (\ref{eq005s}) and (\ref{eq012f})
a specific running of the strong coupling constant
$\alpha_s(T)$ which we show in
Fig.~\ref{su2alpha} (left). Its corresponding running plasma
constant $\Gamma (T)$ is shown in Fig.~\ref{su2alpha} (right). In
Fig.~\ref{su2alpha} we also show two running coupling constants
extracted from lattice measurements in~\cite{kaczmareketal} for
comparison. Our energy density fit suggests lower
values of the running coupling constant.

\begin{figure}[!h]
\begin{center}
\includegraphics[width=0.48\textwidth]{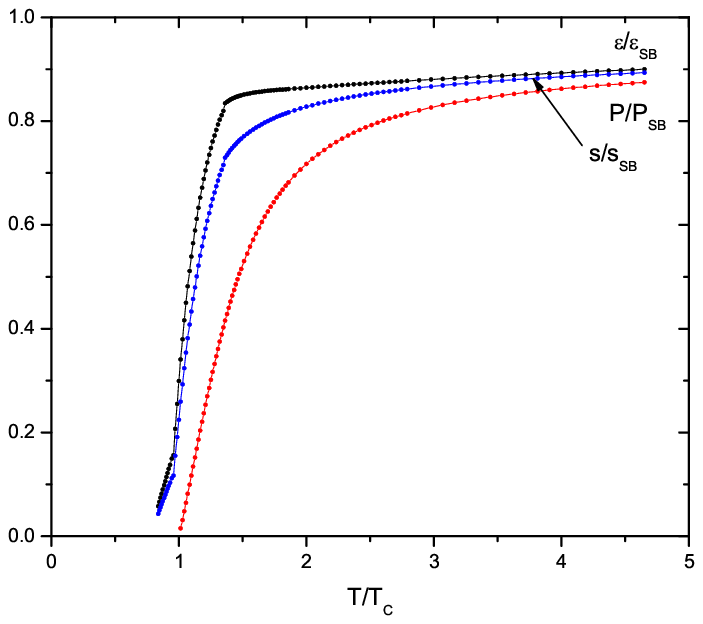}
\includegraphics[width=0.51\textwidth]{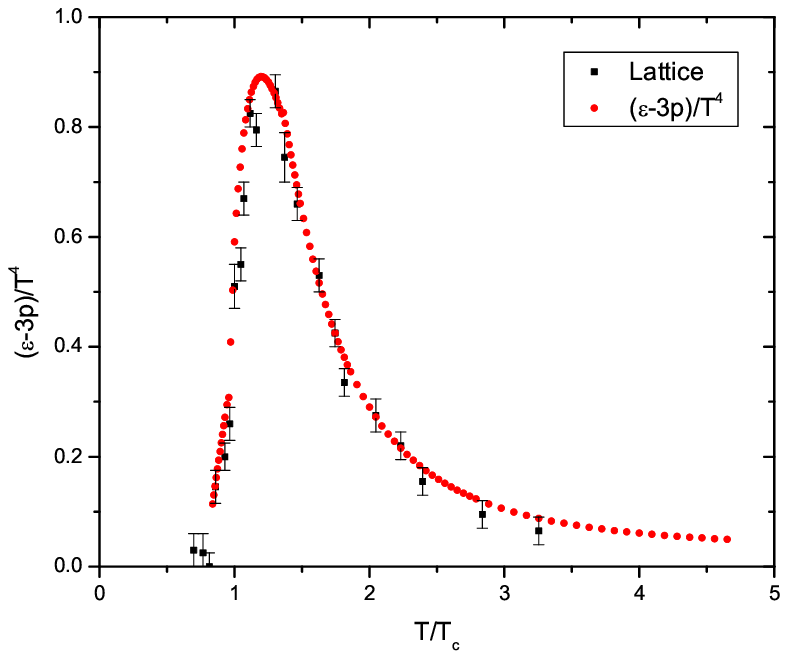}
\end{center}
\caption{Bulk Thermodynamics from the cQGP versus SU(2) lattice. See text.}
\label{e_p_plot_of_su2}
\end{figure}

Since we have fit the energy density to the lattice energy
density, to extract $\alpha_s(T)$, it follows from the Gibbs
relations that all the extensive thermodynamical quantities are
fixed for the SU(2) cQGP. In  Fig.~\ref{e_p_plot_of_su2} (left) we show the
behavior of the energy density, pressure and entropy density
across $T_c$. In  Fig.~\ref{e_p_plot_of_su2} (right) the trace of
the energy momentum tensor is shown versus SU(2) lattice data.

\begin{figure}[!h]
\begin{center}
\includegraphics[width=0.50\textwidth]{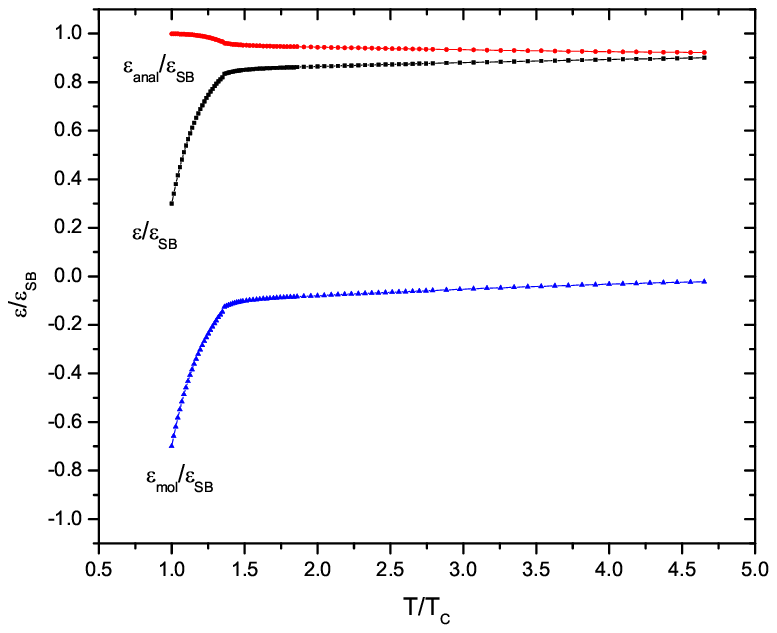}
\includegraphics[width=0.49\textwidth]{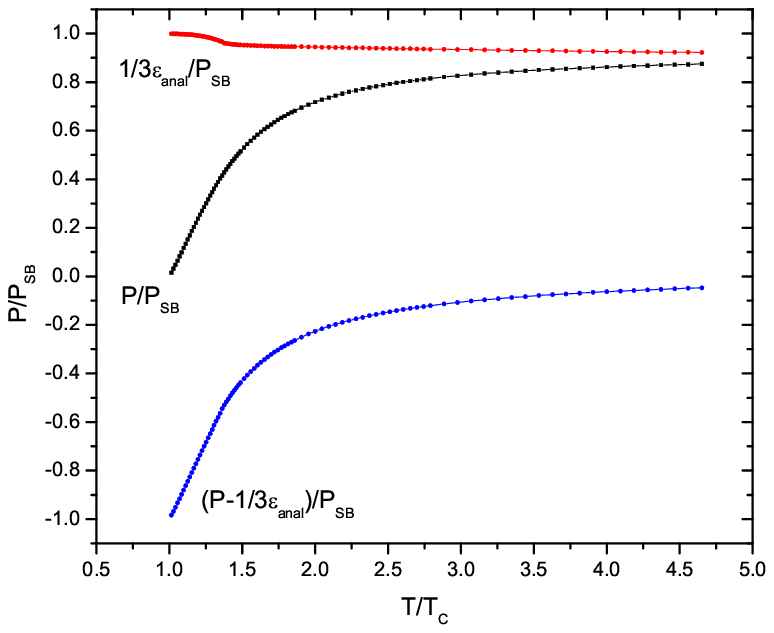}
\end{center}
\begin{center}
\includegraphics[width=0.50\textwidth]{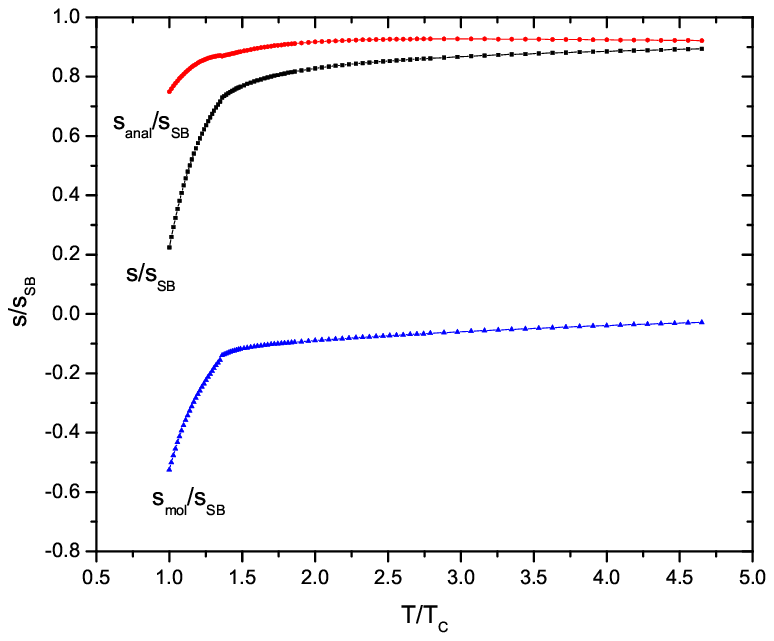}
\includegraphics[width=0.49\textwidth]{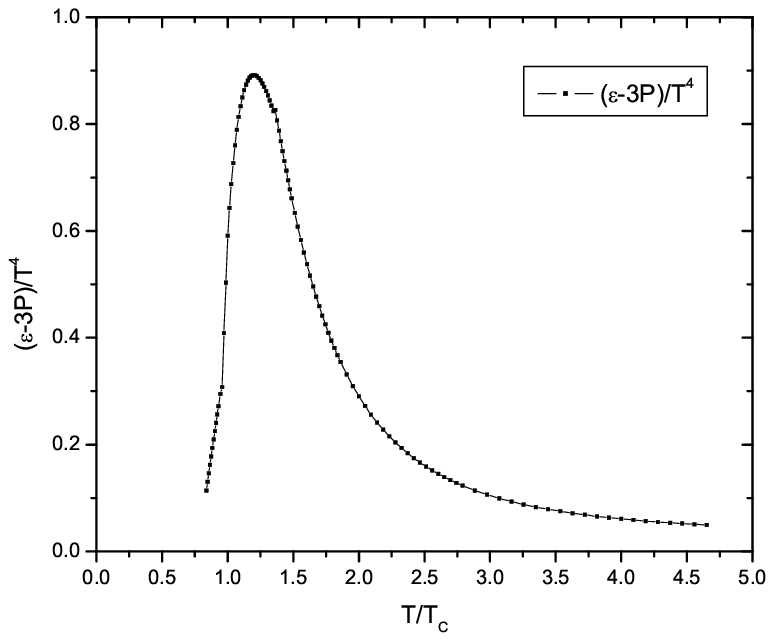}
\end{center}
\caption{Relative contributions in the cQGP bulk thermodynamics. See text.}
\label{separated}
\end{figure}

Our current analysis of the bulk thermodynamics of the cQGP allows
us through the excess energy (\ref{eq005s}) and the Gibbs
relations (\ref{eq010f}) to assess the role of the strongly
coupled component of the cQGP. In Fig.~\ref{separated} we show the
two contributions (loop and molecular) to the bulk thermodynamics
following from the separation

\begin{eqnarray}
& & u_{SU(2),anal}(\Gamma)=\frac{u_{anal}(\Gamma)}
{1+5.5\times10^2\Gamma^{5.4}} \nonumber \\
& & u_{SU(2),mol}(\Gamma)=\frac{5.5\times10^2\Gamma^{5.4}
u_{mol}(\Gamma)} {1+5.5\times10^2\Gamma^{5.4}} \label{eq006s}
\end{eqnarray}
in the energy density. The strongly coupled component of the cQGP
generated by the molecular dynamics simulations contribute
significantly across the transition temperature, say in the range
$(1-2.5)\,T_c$.

\section{Conclusion}

We have constructed the energy density of the cQGP valid for
all values of the plasma parameter $\Gamma$, that interpolates
between the one-loop result at small $\Gamma$ and the molecular
dynamics simulations at large $\Gamma$.  We have used it in
conjunction with the Gibbs relations to derive the Pressure
and entropy of the cQGP.

In quantum QCD $\Gamma$ runs through the QCD coupling constant
at weak coupling. The running at strong coupling is unknown in
general except for some recent lattice simulations
\cite{kaczmareketal}. We have suggested that a fit of our
energy density to the lattice energy density \cite{engelsetal}
allows an extraction of the running coupling that is smaller
than the one suggested by direct lattice simulations
\cite{kaczmareketal}.

We have used the extracted running coupling constant to predict
the entropy density, pressure and energy-momentum trace of the
cQGP. The latter compares well to direct lattice SU(2)
simulations. We have shown that the strongly coupled component of
the cQGP contributes significantly to the bulk thermodynamics
across the transition temperature. We expect transport properties
such as diffusion and viscosity, as well as energy loss to be also
significantly affected in this transition region in the cQGP as we
discuss next~\cite{CHOV}.

\section{Acknowledgments}
This work was supported in part by US-DOE grants DE-FG02-88ER40388
and DE-FG03-97ER4014.

\appendix

\section{Concentration}

\renewcommand{\theequation}{A.\arabic{equation}}
\setcounter{equation}{0}

The bare concentration $c_0=N/V$ which is identified with
the black-body radiation in the cQGP is in general
modified to $c=c_0+\Delta c$ due to interactions. As a
result, the plasma parameter $\Gamma$ (\ref{eq012f}) is
in principle  different from the one used in the text.
The corrected plasma constant is

\begin{eqnarray}
\Gamma_c& &=\Big(\frac{4\pi}{3}c \Big)^{\frac{1}{3}}\beta
C_2\alpha_s(T)=\Big(\frac{4\pi}{3}(c_0 +\triangle c)
\Big)^{\frac{1}{3}}\beta C_2\alpha_s(T)=\Big(
\frac{4\pi}{3}c_0\Big)^{\frac{1}{3}}(1+\frac{\triangle c}{c_0})^{
\frac{1}{3}}\beta C_2\alpha_s(T) \nonumber \\
& & \simeq\Big(\frac{4\pi}{3}c_0\Big)^{\frac{1}{3}} \beta
C_2\alpha_s(T)+\Big(\frac{4\pi}{3}c_0\Big)^{\frac{1}{3}}
\frac{1}{3} \frac{\triangle c}{c_0}\beta
C_2\alpha_s(T)+\mathcal{O}((\frac{\triangle c}{c_0})^2)
\label{eq015f}
\end{eqnarray}
From~\cite{cho&zahed}, the shift in the concentration reads

\begin{equation}
c=c_0+\triangle c=c_0+c_0^{\frac{3}{2}}\pi^{\frac{1}{2}}(N_c^2-1)
(\beta C_2)^{\frac{3}{2}}\alpha_s^{\frac{3}{2}}(T)+\mathcal{O}
(\beta^2) \label{eq016f}
\end{equation}
The corrected plasma constant becomes

\begin{eqnarray}
\Gamma_c & &\simeq\Big(\frac{4\pi}{3}c_0\Big)^{\frac{1}{3}} \beta
C_2\alpha_s(T)+\Big(\frac{4\pi}{3}c_0\Big)^{\frac{1}{3}}
\frac{1}{3} c_0^{\frac{1}{2}}\pi^{\frac{1}{2}}(N_c^2-1)(\beta
C_2)^{\frac{5}{2}}\alpha_s^{\frac{5}{2}}(T) \nonumber \\
& & =\Gamma+\frac{\sqrt{3}}{6}(N_c^2-1)\Gamma^{\frac{5}{2}}
\nonumber \\
& & =\Big(0.244(N_c^2-1)\frac{4\pi}{3}\Big)^{\frac{1}{3}}C_{2}
\alpha_s(T)+\frac{\sqrt{3}}{6}(N_c^2-1)\Big(0.244(N_c^2-1)
\frac{4\pi}{3}\Big)^{\frac{1}{3}\cdot\frac{5}{2}}C_{2}^{\frac{5}{2}}
\alpha_s^{\frac{5}{2}}(T) \nonumber \\
\label{eq017f}
\end{eqnarray}
Inserting (\ref{eq017f}) in the excess energy density yields

\begin{eqnarray}
\frac{U_{loop,ex}(\Gamma)}{\epsilon_{SB}}& & =-\frac{\sqrt{3}}{2}
(N_c^2-1)\Gamma_c^{\frac{3}{2}}+\frac{3}{2}\delta(N_c^2-1)\Gamma_c^2
-\frac{15}{2}\sqrt{3}\delta^2(N_c^2-1)\Gamma_c^{\frac{5}{2}}+\mathcal{O}
(\Gamma_c^3) \nonumber \\
& & \simeq -\frac{\sqrt{3}}{2}(N_c^2-1)\Big(\Gamma+\frac{
\sqrt{3}}{6}(N_c^2-1)\Gamma^{\frac{5}{2}}\Big)^{\frac{3}{2}}+\frac{3}{2}
\delta(N_c^2-1)\Big(\Gamma+\frac{\sqrt{3}}{6}(N_c^2-1)\Gamma^{\frac{5}
{2}}\Big)^2 \nonumber \\
& &-\frac{15}{2}\sqrt{3}\delta^2(N_c^2-1)\Big(\Gamma+\frac{
\sqrt{3}}{6}(N_c^2-1)\Gamma^{\frac{5}{2}}\Big)^{\frac{5}{2}} \nonumber \\
& & \simeq -\frac{\sqrt{3}}{2}
(N_c^2-1)\Gamma^{\frac{3}{2}}+\frac{3}{2}\delta(N_c^2-1)\Gamma^2
-\frac{15}{2}\sqrt{3}\delta^2(N_c^2-1)\Gamma^{\frac{5}{2}}+\mathcal{O}
(\Gamma^3)\label{eq018f}
\end{eqnarray}
which shows that to order $\Gamma^{\frac 52}$ the replacement
$\Gamma_c=\Gamma$ is allowed.

\section{Debye-H\"uckel plus hole (DHH) theory}

\renewcommand{\theequation}{B.\arabic{equation}}
\setcounter{equation}{0}

At strong coupling the Debye-H\"uckel (DH) theory which is
essentially a classical screening theory fails. Debye-H\"uckel
plus Hole (DHH) theory is a way to address DH shortcomings at
strong coupling by building a hole around each charge to account
for the non-penetrability or core in classical
liquids~\cite{nordholm} at higher density or larger $\Gamma$.
As a result, in DHH theory of the cQGP a
color charge density around a test charge is

\begin{equation}
\rho^{\alpha}(r) = \left\{ \begin{array}{ll}
-c\frac{g}{\sqrt{4\pi}}Q^{\alpha}  & (r<\sigma) \nonumber \\
-c\frac{g}{\sqrt{4\pi}}Q^{\alpha}
\frac{\sigma}{r}e^{(-\kappa_{D}(r-\sigma))} & (r\geq \sigma)
\end{array} \right.  \label{eqa01}
\end{equation}
$\sigma$ is the size of the hole, $\alpha$ is a classical color index $(1,..,N_c^2-1)$,
$\beta=1/T$  and $\kappa_D$ is the Debye momentum

\begin{equation}
\kappa_D^2=\frac{g^2}{N_{c}^2-1}c \beta
\sum_{\alpha}^{N_c^2-1}{Q^{\alpha}}^2\,\,. \label{eqa02}
\end{equation}
The negative sign in (\ref{eqa01}) reflects on the screening, with
the Debye cloud left unchanged outside $\sigma$. The hole size
$\sigma$ is fixed by demanding that each test particle is
completely screened through

\begin{equation}
\int_{0}^{\infty}dr 4\pi r^2 \rho^{\alpha}(r)=-
\frac{g}{\sqrt{4\pi}}Q^{\alpha} \label{eqa03}
\end{equation}
This condition, fixes the size of the hole

\begin{equation}
\sigma=\frac{1}{\kappa_{D}}\Big((1+\frac{3\kappa_{D}^3} {4\pi c}
)^{\frac{1}{3}}-1\Big) \label{eqa04}
\end{equation}
In terms of

\begin{equation}
\Gamma=\frac{g^2}{4\pi}\frac{C_{2}}{T a_{WS}} \label{eqa05}
\end{equation}
the hole radius is

\begin{equation}
\sigma=\frac{1}{\kappa_{D}}\Big((1+(3\Gamma)
^{\frac{3}{2}})^{\frac{1}{3}} -1\Big)\label{eqa06}
\end{equation}
after fixing the Wigner-Seitz radius $a_{WS}$ through $c_0 (4\pi
a_{WS}^3/3)=1$. Again we have set $c=c_0$. From (\ref{eqa06}) it
follows that the hole size is smaller the higher the density or
temperature.


\begin{thebibliography}{1}

\bibitem{ichimaru} S.Ichimaru, Rev. Mod. Phys. \textbf{54}, 1017 (1982)
\bibitem{hansen&mcdonald} J.Hansen and I.Mcdonald, \textit{Theory of Simple Liquids} (Academic Press, 2006)
\bibitem{hansenetal} J.Hansen, I.McDonald and E.Pollock, Phys. Rev. A \textbf{11}, 1025 (1975)
\bibitem{borisetal} B.A.Gelman, E.V.Shuryak and I.Zahed, Phys. Rev. C \textbf{74}, 044908 (2006)
[arXiv:nucl-th/0601029]
\bibitem{cho&zahed} S.Cho and I.Zahed [arXiv:0812.1736]
\bibitem{borisetal2} B.A.Gelman, E.V.Shuryak and I.Zahed, Phys. Rev. C \textbf{74}, 044909 (2006)
[arXiv:nucl-th/0605046]
\bibitem{ichimaru2} S.Ichimaru, \textit{Statistical Plasma Physics Vol II:Condensed Plasmas} (Westview Press, 2004)
\bibitem{bannur} V.Bannur, Eur. Phys. J. C \textbf{11}, 169 (1999)
[arXiv:hep-ph/9811397]
\bibitem{nordholm} S.Nordholm, Chem. Phys. Lett. \textbf{105}, 302 (1984)
\bibitem{engelsetal} J.Engels, F.Karsch and H.Satz, Phys. Lett. B \textbf{101}, 89 (1981)
\bibitem{kaczmareketal} O.Kaczmarek, F.Karsch, F.Zantow and P.Petreczky, Phys. Rev. D \textbf{70}, 070405
(2004) [arXiv:hep-lat/0406036]
\bibitem{CHOV} S. Cho and I. Zahed, in preparation.





\end{thebibliography}
\end{document}